 \documentclass{emulateapj}

\newcommand{\QSO}     {SDSS\,J2222$+$2745}
\newcommand{\QSOLone}    {SDSS\,J1004$+$4112}
\newcommand{\QSOLtwo}    {SDSS\,J1029$+$2623}
\newcommand{\etal}    {{\it et al.~}}

\newcommand{\ABdisp} {47.7 \pm 6.0}
\newcommand{\shortABdisp} {47.7}
\newcommand{\ABdispoffset} {0.340 \pm 0.007}
\newcommand{\ABchisq} {42.0^{+1.1}_{-0.1}}
\newcommand{\ABchisqoffset} {0.345^{+0.006}_{-0.005}}

\newcommand{\ACdisp} {722 \pm 24}
\newcommand{\ACdispoffset} {0.483 \pm 0.012}
\newcommand{\ACchisq} {717.4 \pm 0.5}
\newcommand{\ACshort} {722}
\newcommand{\ACchisqoffset} {0.504 \pm 0.013}
\newcommand{\ACshortoffset} {0.483}

\shorttitle{Time Delays for \QSO }
\slugcomment{Accepted for publication in ApJ}
\shortauthors{Dahle \etal}

\begin{document}
\title{Time Delay Measurements for the Cluster-lensed Sextuple Quasar
  \QSO . \footnote{Based on observations made with the Nordic Optical
    Telescope, operated on the island of La Palma jointly by Denmark,
    Finland, Iceland, Norway, and Sweden, in the Spanish Observatorio
    del Roque de los Muchachos of the Instituto de Astrofisica de
    Canarias, and including observations obtained at the Gemini
    Observatory, which is operated by the Association of Universities
    for Research in Astronomy, Inc., under a cooperative agreement
    with the NSF on behalf of the Gemini partnership: the National
    Science Foundation (United States), the National Research Council
    (Canada), CONICYT (Chile), the Australian Research Council
    (Australia), Minist\'{e}rio da Ci\^{e}ncia, Tecnologia e
    Inova\c{c}\~{a}o (Brazil) and Ministerio de Ciencia,
    Tecnolog\'{i}a e Innovaci\'{o}n Productiva (Argentina).  }}
\author{ H.~Dahle\altaffilmark{1}, M.~D.~Gladders\altaffilmark{2,3},
  K.~Sharon\altaffilmark{4}, M.~B.~Bayliss\altaffilmark{5,6},
  J.~R.~Rigby\altaffilmark{7}}

\email{hdahle@astro.uio.no} 

\altaffiltext{1} {Institute of Theoretical Astrophysics, University of Oslo, P. O. Box 1029, Blindern, N-0315 Oslo, Norway} 
\altaffiltext{2} {Department of Astronomy \& Astrophysics, The University of Chicago, 5640 S.\ Ellis Avenue, Chicago, IL 60637, USA} 
\altaffiltext{3} {Kavli Institute for Cosmological Physics at the University of Chicago}
\altaffiltext{4} {Department of Astronomy, University of Michigan, 500 Church Street, Ann Arbor, MI 48109, USA} 
\altaffiltext{5} {Harvard-Smithsonian Center for Astrophysics, 60 Garden Street, Cambridge, MA 02138, USA} 
\altaffiltext{6} {Department of Physics, Harvard University, 17 Oxford Street, Cambridge, MA 02138} 
\altaffiltext{7} {Observational Cosmology Lab, NASA Goddard Space Flight Center, Greenbelt, MD 20771, USA}

\begin{abstract}

We report first results from an ongoing monitoring campaign to measure time delays between the six images of the quasar \QSO \,, gravitationally 
lensed by a galaxy cluster. The time delay between A and B, the two most highly magnified images, is measured to be $\tau_{\rm AB} = \ABdisp$\,days 
(95\% confidence interval), consistent with previous model predictions for this lens system.  The strong intrinsic variability of the quasar also allows us to derive a time delay value of $\tau_{\rm CA} = \ACdisp$\,days between image C and A, in spite of modest overlap between their light curves 
in the current data set. 
Image C, which is predicted to lead all the other lensed quasar images, has undergone a sharp, 
monotonic flux increase of 60-75\% during 2014.  A corresponding brightening is firmly predicted to occur in images A and B during 2016.   
The amplitude of this rise indicates that time delays involving all six known images in this system, including those of the demagnified central 
images D-F, will be obtainable from further ground-based monitoring of this system during the next few years. 
     
\end{abstract}


\section{Introduction} 
\label{sec:intro}

Quasars gravitationally lensed into multiple images by foreground clusters of galaxies are exceptionally rare. The two first reported cases were
\QSOLone \, (Inada et al.\ 2003) and \QSOLtwo \, (Inada et al.\ 2006). A third case, \QSO \,, was discovered by Dahle et al.\ (2013), displaying     
six images (of which five were spectroscopically confirmed) of a quasar at redshift $z_s=2.82$, with a maximum separation of $15\farcs 1$, lensed 
by a foreground cluster at $z_s=0.49$. In contrast, the Master Lens Database of known gravitational lens systems (Moustakas et al.\  2012) lists 
$\sim 120$ known galaxy-scale quasar lenses. 

Refsdal (1964) showed how the relative time delays measured in multiply lensed quasar images can be used to measure the Hubble parameter $H_0$. In addition, time delay measurements can probe the underlying cosmological model (Refsdal 1966), yielding dark energy constraints which are complementary to other such probes (Linder 2011; Treu et al.\ 2013; Sereno \& Paraficz 2014). Time delay measurements of individual systems are subject to a "cosmic variance" 
of several percent, caused by the mass fluctuations of intervening structure along the line of sight (e.g., Bar-Kana 1996; Wambsganss et al.\ 2005). 
Aside from the cosmographical applications of time delays, they also probe the surface density distribution of the lens, over the range 
of radii at which images are seen and time delays are measured.  A unique property of \QSO \, is the availability, at small clustercentric radii, 
of three fainter images (D-F)  which are predicted to be demagnified, according to the six-image model presented by Dahle et al.\ (2013). 
These three images are still sufficiently bright to have their time delays measured from ground-based data, given the 
level of quasar variability reported in this paper. 

The longest time delays produced by cluster-scale lenses can be of order several years (Fohlmeister et al.\ 2008; 2013), rather than weeks or months in the case of galaxy-scale lenses. 
With typical monitoring cadence of a few days to a few weeks, and given the measured trend of quasar variability amplitude to increase as a function of time scale 
(at least up to time scales of several years; e.g., Vanden Berk et al.\ 2004), cluster-scale lenses may yield smaller fractional uncertainties on the time delays, 
making cosmic variance-limited time delay measurements easier to obtain than for galaxy-scale lenses. 
A further advantage of cluster-scale lenses is the availability of additional constraints on the lensing mass distribution from other multiply lensed background galaxies 
as exemplified by the HST-based studies of \QSOLone \, (Sharon et al.\ 2005; Oguri 2010) and \QSOLtwo \, (Oguri et al.\ 2013).     

Measuring the time delays in wide-separation cluster-lensed quasars also enables three-dimensional studies of outflowing winds within the quasar itself by probing multiple sightlines, as shown by  Misawa et al.\  (2014). In this case, the time delays need to be known in order to disentangle time  
variability of absorption profiles in the quasar image spectra from genuine differences between the sightlines.  
 
Finally, knowing the time delay is crucial for advance planning of monitoring with increased frequency in anticipation of predicted strong flux changes, 
e.g., to measure the time delays of the faintest images in the system or to conduct reverberation mapping studies to measure the mass of the black hole that powers the quasar (Blandford \& McKee 1982; Peterson 1993, 2014). 
     
In this paper, we summarise results obtained from the first three years of an ongoing monitoring campaign of \QSO \,.    
In \S~\ref{sec:data} we present the data set available so far, which is used to derive a value of the time delay $\tau_{AB}$ in \S~\ref{sec:Delta_t} and a value of the 
longer delay $\tau_{CA}$ in \S~\ref{sec:longer}.  
In \S~\ref{sec:discuss}, we summarise and discuss the results, including a strong rise in the flux of image C during 2014, and the resulting prospects for future studies of this system.        
 
\section{Data} 
\label{sec:data}

A monitoring campaign to measure time delays in \QSO \, was initiated at the 2.56m Nordic Optical Telescope (NOT) in October, 2012. 
The results in this paper are based on data from this ongoing campaign, with the most recent data obtained in August, 2015.  
The NOT monitoring program makes use of the Andalucia Faint Object Spectrograph and Camera (ALFOSC) instrument, 
which has a $2048^2$ pixel CCD detector with 32-bit intensity resolution and a pixel scale of $0\farcs 188\,{\rm pixel}^{-1}$. 
At each monitoring epoch, a set of either $3\times 300$s or $3\times 600$s exposures were obtained.   

The annual visibility period of \QSO \, extends from late April until early January. We chose the SDSS $g$-band for our monitoring campaign, in order to maximise the signal-to-noise of the quasar images, relative to the galaxies in the $z=0.49$ lensing cluster on which the central D-F images 
are superposed. This band is well centered on the prominent Lyman-alpha line in the spectrum of the $z=2.82$ quasar, 
at an observed wavelength of 4640~{\AA}. 

In addition to the ALFOSC data, we include $g$-band data obtained as
part of the original discovery of this lens (for further details, see
Dahle et al.\ 2013), using the MOSaic CAmera (MOSCA) at NOT. MOSCA is
a $2\times 2$ mosaic of four $2048 \times 2048$ CCDs used in $2\times
2$ binned mode, yielding a pixel scale of $0\farcs 217\,{\rm pixel}^{-1}$. MOSCA $g$-band imaging was obtained for three separate
epochs, on 2011 Sep.\ 24.93, 2012 Sep.\ 12.96, and 2012 Sep.\ 15.98
UT, respectively. We also include the original SDSS DR8 $g$-band
photometry for this system, obtained on 2009 October 18.19
UT. Finally, this paper also includes recently acquired Gemini imaging
of \QSO \, (taken under program GN-2015A-FT-16); these few images
represent the beginning of a new observational campaign to measure
time delays for the much fainter central D-F images. Those data were
acquired in $g$-band with Gemini North, using the GMOS instrument in
imaging mode, on 2015 May\ 25.54, 2015 May\ 26.52, and 2015
July\ 25.48 UT. 

\begin{figure}
\begin{center}
\includegraphics[angle=0,scale=.4]{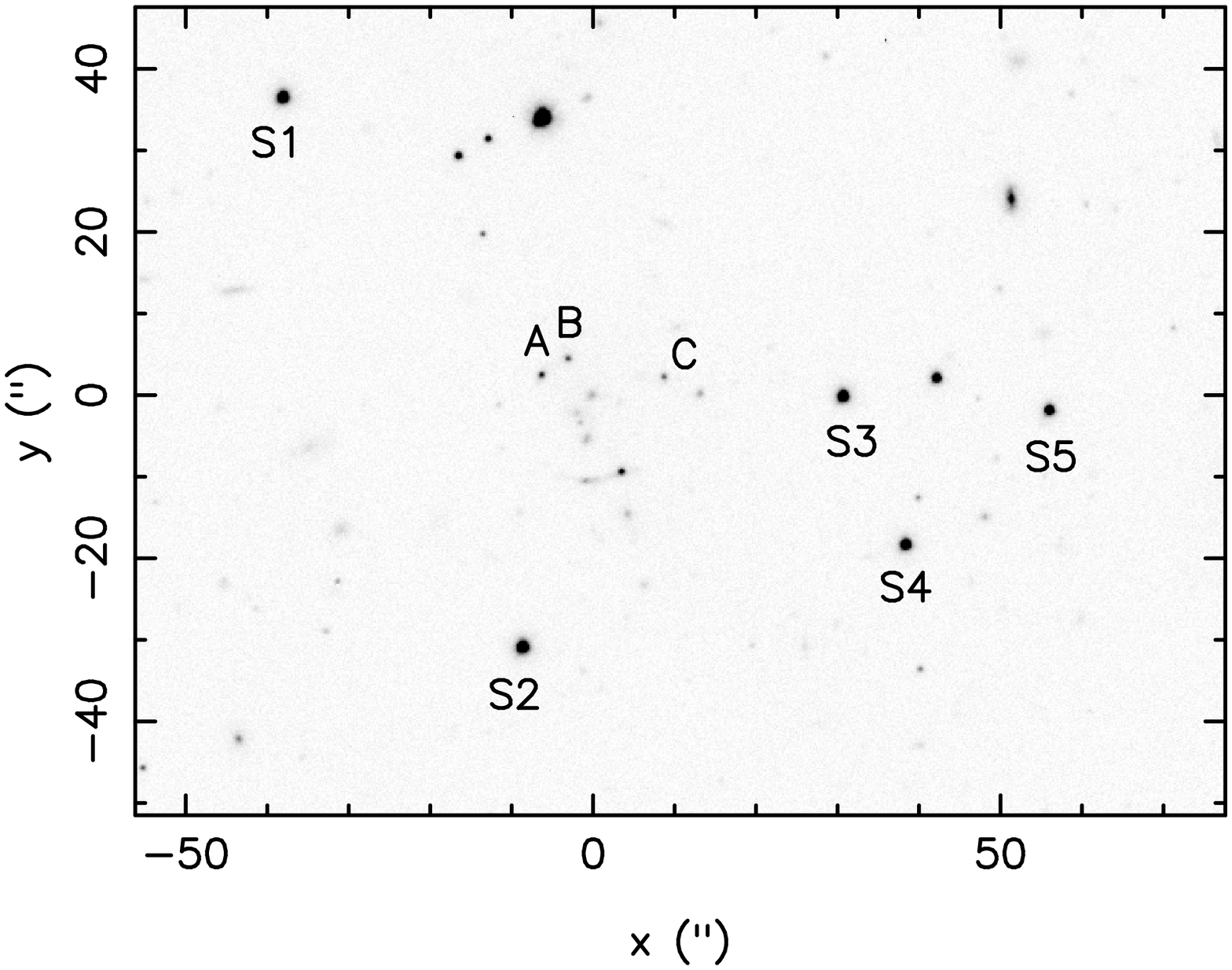}
\end{center}
\caption[Comparison stars.]
{A single $600$s ALFOSC exposure, with the locations of the five comparison stars S1-S5 indicated. }
\label{fig:S1S5}
\end{figure}

In this paper, we report photometric measurements of the three
brightest (A-C) images of \QSO \,. Unlike the fainter (D-F) images,
these are all well separated from foreground objects and their fluxes
can be reliably measured without modelling and removal of additional
sources. In order to calibrate the observed fluxes, we use a set of 5
reference stars (S1-S5; see Fig.~\ref{fig:S1S5}). In order to minimise
the effect of spatial variations of the PSF, these are chosen within a
radius $1\arcmin$ from the quasar images. In addition, the reference
stars have blue colors similar to that of the quasar and do not have
any nearby objects which may affect the flux estimate.  Even during
the best seeing (FWHM $\sim 0\farcs5$), these stars have measured peak
fluxes $\lesssim 80$\,kADU, well within the linear regime of the ALFOSC
detector, which extends up to $\sim 350-400$\,kADU. The photometric
measurements in our data were all based on aperture photometry of the
quasar images and reference stars, using the same aperture.  An
aperture correction was implicitly made by setting the magnitude zero
point of each exposure such that the mean magnitude value of the
ensemble of 5 reference stars equals their mean $g$-band magnitude
value in the SDSS ($g = 19.092$).  The accuracy of the time delay
measurements reported in this paper is not dependent on a highly
accurate absolute flux calibration, as long as the reference stars do
not vary significantly in brightness.  The validity of using S1-S5 as
flux reference sources was tested by measuring the rms fluctuation in
the magnitude offset of each star from the ensemble average, over the
time span covered by our ALFOSC observations.  The rms value of these
fluctuations range between 0.0061 and 0.0095 magnitude. This is two
orders of magnitude smaller than the amplitude of the flux variations
measured in \QSO \,, as described below.

The ALFOSC imaging data were de-biased and flat fielded using standard
techniques.  Flat-fielding was typically performed using the median of
a series of flat field frames of the twilight sky obtained during the
same night as the photometric data. As the ALFOSC detector is
cosmetically clean, the three individual exposures for each monitoring
epoch were treated individually in our analysis rather than combined
into a single frame.

The Gemini GMOS imaging data were de-biased and flat fielded, again
using standard techniques, with appropriate and temporally adjacent
master bias and flat field frames taken from the Gemini Science
Archive. Each Gemini observation is the sum of six dithered 300 second
integrations, stacked to eliminate cosmic rays. Care was taken to
avoid improperly rejecting pixels in the cores of bright stars when
stacking the dithered frames. Though images A-C are well measured in
each GMOS sub-image, the thick red-sensitive CCDs in use on GMOS-North are
also excellent charged particle detectors, and the most robust
measurements thus come from stacked images with good cosmic ray
rejection.

The fluxes of \QSO \, images A-C and reference stars S1-S5 were all
measured within a fixed aperture of diameter $1\farcs 13$, for the
ALFOSC, MOSCA and GMOS data. The choice of aperture scale was
motivated by containing most of the flux while avoiding contamination
from nearby sources.  The effect on the time delay measurements of
choosing an aperture a factor 2 smaller was tested, and found to
produce changes in the time delay estimate within the uncertainties
quoted below.  The photometric uncertainty was
determined by measuring the photometric scatter corresponding to the
three individual ALFOSC exposures and fitting the result as a function of the
signal-to-noise of the quasar images. 

\begin{figure}
\begin{center}
\includegraphics[angle=-90,scale=.30]{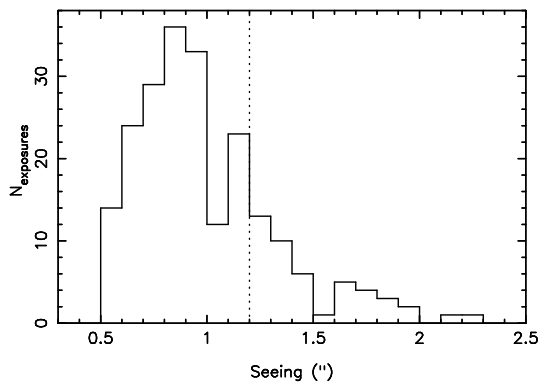}
\end{center}
\caption[Seeing in ALFOSC images.]
{Histogram of seeing values measured in individual ALFOSC exposures. The dotted vertical line corresponds to the upper cutoff value for data used in our analysis.}
\label{fig:seeing}
\end{figure}

Figure~\ref{fig:seeing} shows a histogram of seeing values measured in the individual ALFOSC monitoring exposures. 
In our further analysis, we excluded data points obtained when the FWHM seeing was worse than $1\farcs 20$, and we also excluded 
data where the sky background level was very high, e.g., caused by moonlight or twilight. This left 42 distinct epochs of ALFOSC monitoring 
observations (not counting multiple exposures within the same night), with an average time span between observations of 14 days, during the annual visibility period of this target. The photometric measurements used in our further analysis are listed in Table~\ref{tab:photometrytable}.      

\begin{figure}
\plottwo{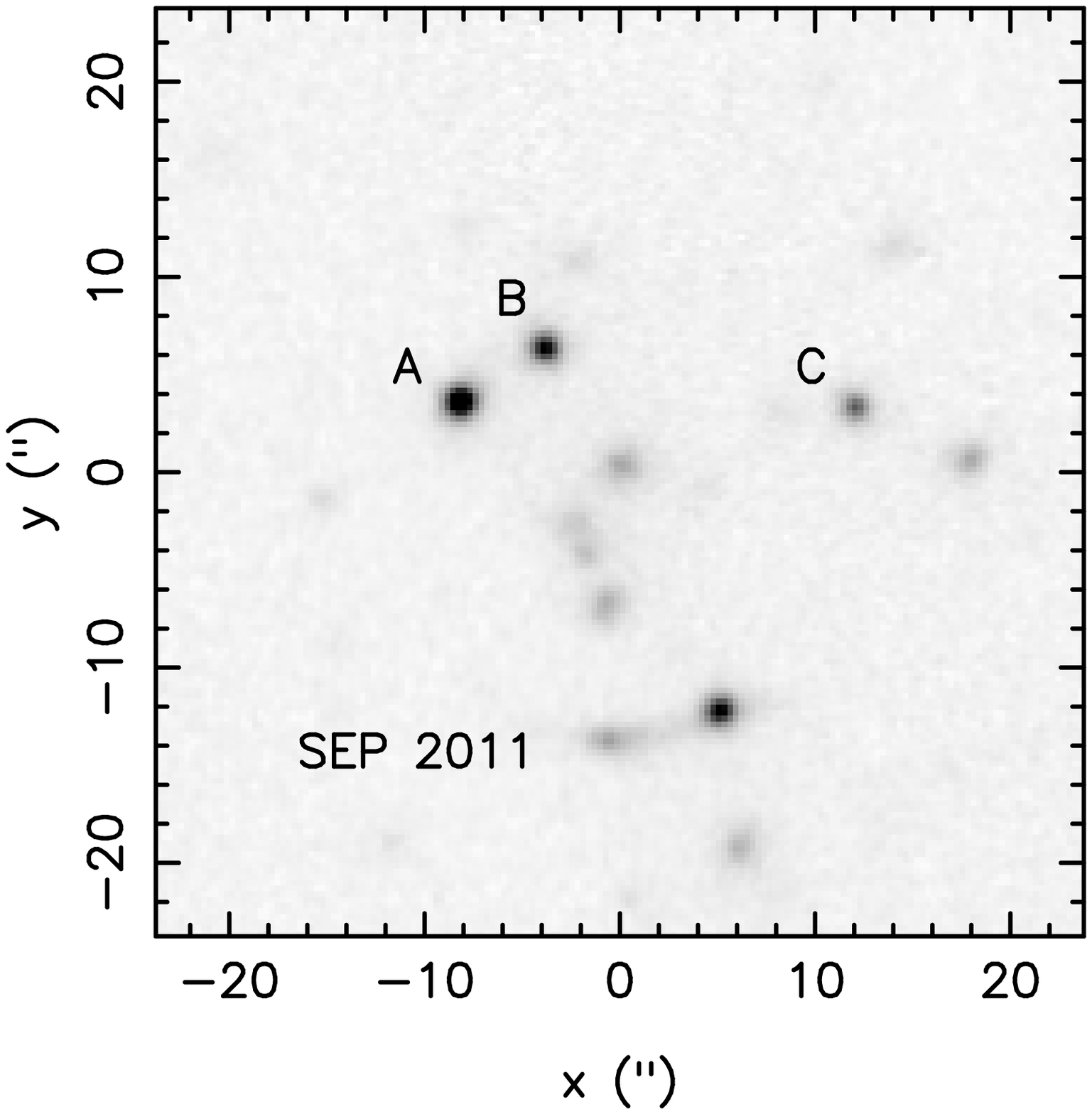}{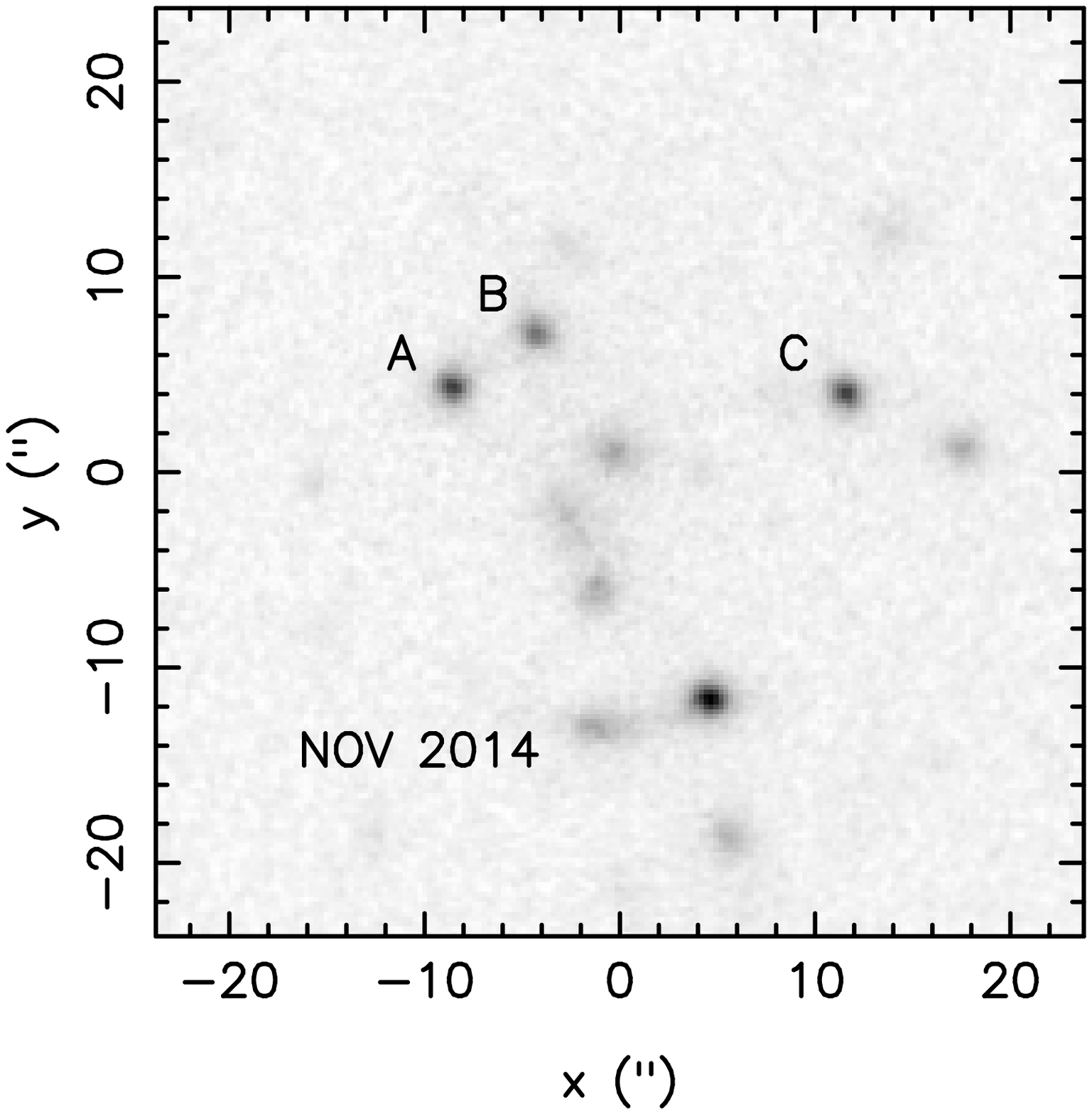}
\caption[.]  {Left Panel: MOSCA image from September 2011. Right Panel: ALFOSC image from November 2014, illustrating the flux variation of the C image relative to the A and B images of the quasar during our monitoring of \QSO . 
 }
\label{fig:fluxvar}
\end{figure}

\begin{figure}
\includegraphics[angle=-90,scale=.36]{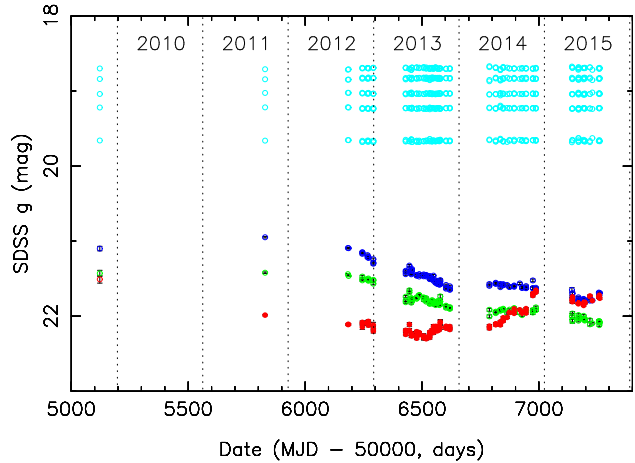}
\caption[.]  {Light curves of the quasar images A (blue symbols), B (green), and C (red).  The light blue points in the upper part 
of the image represent the photometric measurements of the comparison stars S1-S5.}
\label{fig:lightcurves}
\end{figure}

Significant flux variability of the three brightest images is evident even from a cursory inspection of the imaging data: Figure~\ref{fig:fluxvar} 
qualitatively illustrates the flux changes over a 3.2\,yr time span:  In September 2011, the A and B images 
were both obviously brighter than the C image. By November 2014, the A and B components 
had both faded while approximately preserving the A:B flux ratio, while the C image had brightened to the point where it is clearly brighter 
than the B component and rivals the A component in magnitude. A quantitative representation of these trends is provided by 
the light curves of the three brightest components, shown in Figure~\ref{fig:lightcurves}.   

\section{The A-B time delay} 
\label{sec:Delta_t}

Model predictions for the time delays between the six observed quasar images were reported in the work by 
Dahle et al.\ (2013). Being based on limited constraints from ground-based imaging (e.g., using only a single family 
of multiply lensed images, of the quasar itself), these predictions all have large uncertainty intervals of several hundred days.     

The model prediction for the time delay between the A and B images reported by Dahle et al.\ (2013) is  
$\tau_{\rm AB} = 112^{+225}_{-158}$ days, where positive values imply that image A is leading image B.  
The time delay between A and C is predicted to be $\tau_{\rm CA} = 1256^{+485}_{-411}$\,days. 
Given the much shorter predicted time delay between A and B, we would expect the 
quasar variability light curves of these two images to overlap over a large fraction of the 
time interval covered by our photometric monitoring. Hence, we focus initially on determining the value of $\tau_{\rm AB}$. 

While a large number of different methods for deriving the time delays of light curves of strong lenses 
have recently been proposed and tested (Liao et al.\ 2015), we focus here on two traditional methods for time delay measurements, detailed below. 
However, even prior to the implementation of these methods, an initial estimate of $\tau_{\rm AB}$ of $\sim 40-50$ days was made
by shifting and visually matching the light curves. 
Finding a plausible match was facilitated by the obvious overall variability trends seen in the A and B light curves, showing continuous   
declines over $\sim 1.5$\, yr, followed by one observing season of nearly constant fluxes during 2014.       
Below, we use two different statistical methods to estimate $\tau_{\rm AB}$ in a more rigorous manner.

\subsection{ $\chi^2$ minimization } 
\label{sec:chisquared}

This method is based on shifting the two observed light curves $A(t_i)$ and $B(t_i)$, cross-correlating each of them with a
linearly interpolated light curve for the other image, ($b(t)$, and $a(t)$, respectively), and finding the values of $\tau$ and 
the magnitude offset $\Delta m$ which minimise the statistic    

   \begin{eqnarray} 
     \frac{\chi^2 (\tau)}{N_{\rm dof} (\tau ) }  = \frac{1}{2 N_{\rm dof} (\tau ) } \sum\limits_{i}^{N (\tau)} \frac{[A (t_i) - b(t_i - \tau ) + \Delta m]^2}{\sigma^2_{A,i} + \sigma^2_{b,t}}       \nonumber      \\
      +  \frac{1}{2 N_{\rm dof} (\tau ) } \sum\limits_{i}^{N (\tau)} \frac{[a (t_i + \tau) - B(t_i ) + \Delta m]^2}{\sigma^2_{a,t} + \sigma^2_{B,i}} \,\cdot      
\end{eqnarray}

\begin{figure}
\begin{center}
\includegraphics[angle=-90,scale=.35]{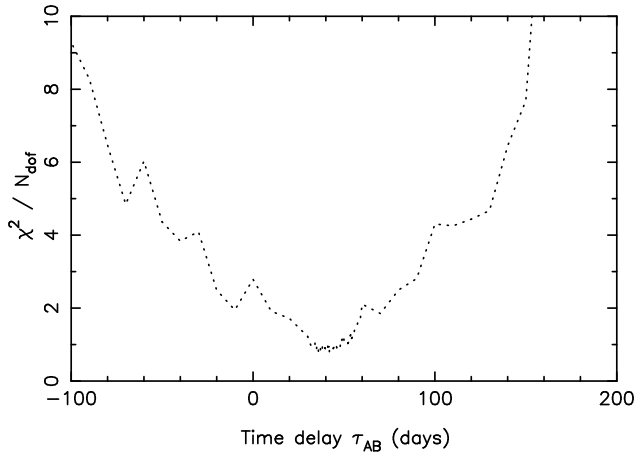}
\end{center}
\caption[Least squares minimisation.]
{The result of a $\chi^2$ minimization obtained by shifting and matching the observed light curves for A and B measured during the period September 2012 - August 2015. }
\label{fig:leastsquares}
\end{figure}

Here, $\sigma_{A,i}$ and $\sigma_{B, i}$ are errors in the observed magnitudes and  $\sigma_{a,t}$ and $\sigma_{b, t}$ are errors in the interpolated magnitudes. We used data from all four seasons of systematic monitoring and assumed that $\Delta m_{AB}$ did not change
between seasons (such changes might be produced by microlensing). 
We did not interpolate between any set of points that spanned the gap between different seasons. 
The minimum value of $\frac{\chi^2 (\tau)}{N_{\rm dof} (\tau )}$ as a function of the time delay is shown in Figure~\ref{fig:leastsquares}.  
We find a minimum value for a time delay $\tau_{\rm AB} = \ABchisq$ days and 
magnitude offset $\Delta m_{\rm AB} = \ABchisqoffset$ (with uncertainties corresponding to $\Delta \chi^2 = 4$). 

\subsection{Dispersion method } 
\label{sec:dispersionmethod}

A shortcoming of the $\chi^2$ minimisation method is the assumption of linear flux variations between 
measured points in the light curve. A more statistically robust alternative method for time delay estimates 
was proposed by Pelt et al.\ (1994; 1996). 

The method constructs a combined data set $C_k (t_k)$ from the two series $A_i$ (with $N_A$ measurements) and 
$B_j$ (with $N_B$ measurements), sampling the quasar light curve:    
  
\begin{equation} 
C(t_k) = \left\{
	\begin{array}{ll}
		A_i & ~~~~t_k = t_i  \\
		 B_j - \Delta m & ~~~~t_k = t_j + \tau , 
		 \end{array}
		 \right.
\end{equation} 

where $k = 1,...,N$ and $N = N_A + N_B$. 

From this combined data set, the dispersion spectrum is defined by:   

\begin{equation}
D^2(\tau ) = \min\limits_{\Delta m} \frac{\sum\limits_{k=1}^{N-1} S_k W_k G_k (C_{k+1} - C_k)^2}{2 \sum\limits_{k=1}^{N-1} S_k W_k G_k},    
\end{equation} 

where $G_k = 1$ only if the measurements are from different images and $G_k = 0$ otherwise, 
$W_k = ( \sigma_k^2  + \sigma_{k+1}^2 )^{-1}$ 
are statistical weights, and   

\begin{equation} 
S_k = \left\{
	\begin{array}{ll}
		1 ~~~~& \left| t_{k+1} - t_k \right| \leq \delta   \\
		 0 ~~~~& \left| t_{k+1} - t_k \right| > \delta .   \\
		 \end{array}
		 \right.
\end{equation} 

\begin{figure}
\begin{center}
\includegraphics[angle=-90,scale=.35]{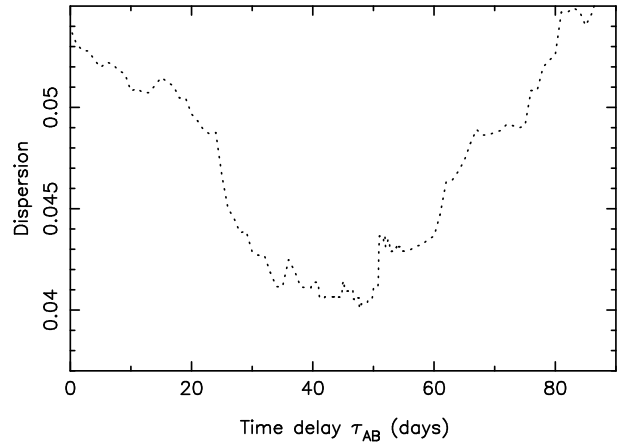}
\end{center}
\caption[Dispersion spectra.]
{Dispersion calculated as a function of the shift $\tau_{\rm AB}$ between the two light curves. }
\label{fig:dispersion}
\end{figure}

We used $\delta = 25$\,days as the decorrelation time scale, 
producing the dispersion spectrum shown in Figure~\ref{fig:dispersion}.  The minimum value is reached for $\tau_{\rm AB} = \ABdisp$\,days and $\Delta m_{\rm AB} = \ABdispoffset$ (95\% confidence intervals). The result has little sensitivity to the choice of $\delta$; 
choosing a number twice or half the value quoted above changes the best fit values by a few percent, well within the uncertainty intervals
quoted below.

To estimate the uncertainties, we used a bootstrap procedure. The light curve was smoothed using a running 9-point median filter and 
the residuals of our data points were calculated with respect to the smoothed light curve. A set of 1000 simulated light curves were then generated 
by randomly re-shuffling the residuals, and the errors were estimated from the spread of values of $\tau_{\rm AB}$ from the 1000 dispersion spectra.    
A combined A+B light curve $C(t_k)$ is represented by the union of the red and blue points plotted in Figure~\ref{fig:addC}. In this figure, 
the observed light curve of the B component has been shifted by the $\tau_{\rm AB}$ and $\Delta m_{\rm AB}$ values derived using the dispersion method. 

\section{The A-C time delay} 
\label{sec:longer}

\begin{figure}
\begin{center}
\includegraphics[angle=-90,scale=.35]{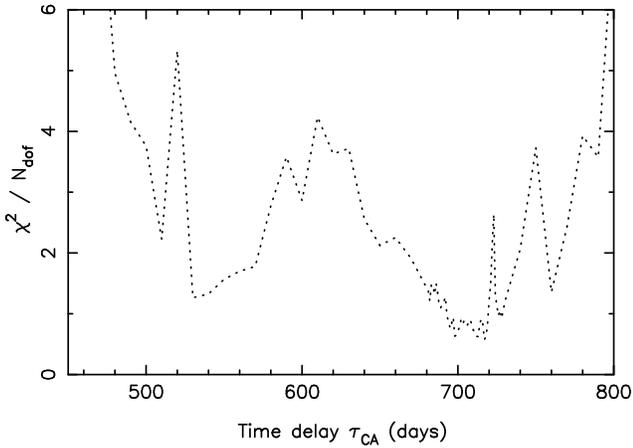}
\end{center}
\caption[Least squares minimisation for C.]
{The result of a $\chi^2$ minimization obtained by shifting and matching the observed light curve of C to the combined A+B light curve. }
\label{fig:leastsquaresC}
\end{figure}

Having estimated $\tau_{\rm AB}$, we next consider the time delay between the combined $AB$ light curve
and image C. For this purpose, we also include the 2009 photometry from SDSS DR8.  All acceptable lens models of this system predict that 
image C leads all the other images, so we do not consider time delays  $\tau_{\rm CA} < 0$. 
A $\chi^2$ minimisation using the procedure in \S~\ref{sec:chisquared} yields a minimum at time delay $\tau_{\rm CA} = \ACchisq$\,days, 
with uncertainties corresponding to $\Delta \chi^2 = 4$. The magnitude offset is constrained as $\Delta m_{\rm AC} = \ACchisqoffset$. 
The minimum value of $\frac{\chi^2 (\tau)}{N_{\rm dof} (\tau )}$ as a function of the time delay is shown in Figure~\ref{fig:leastsquaresC}.  
No meaningful calculation could be done for time delay values greater than 830\,days, since there would so far be minimal or no overlap 
between the monitoring light curves for such long time delays. 

\begin{figure}
\begin{center}
\includegraphics[angle=-90,scale=.35]{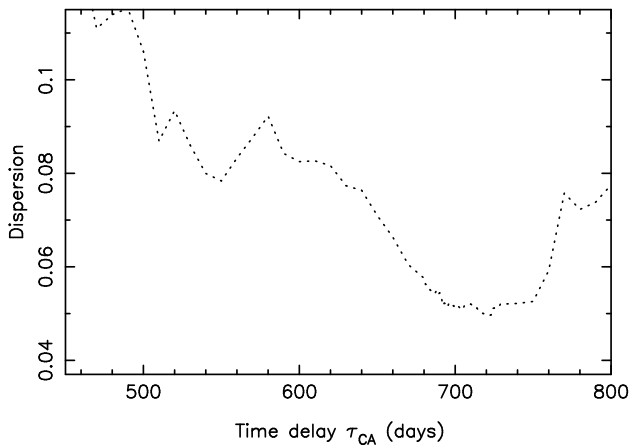}
\end{center}
\caption[Dispersion spectrum for C.]
{Dispersion calculated as a function of the shift $\tau_{CA}$ between the observed light curve of image C and the combined A+B light curve.}  
\label{fig:dispersionC}
\end{figure}

We also estimated  $\tau_{\rm CA}$ using the dispersion method described in \S~\ref{sec:dispersionmethod}. This produced the dispersion 
spectrum shown in Figure~\ref{fig:dispersionC}, yielding a best-fit time delay of  $\tau_{\rm CA} = \ACdisp$\,days and magnitude offset $\Delta m_{\rm AC} = \ACdispoffset$ (95\% confidence intervals). These values are consistent with the values produced by the $\chi^2$ minimisation method. 
Since the dispersion method does not rely on assumptions about the quasar flux variation between monitoring epochs, we consider the values and 
uncertainties produced by this method to be more reliable than those produced by interpolation and $\chi^2$ minimisation. Hence, we adopt 
the time delay $\tau_{\rm CA} = \ACdisp$\,days in our further discussion. 

\begin{figure*}
\includegraphics[angle=-90,scale=.775]{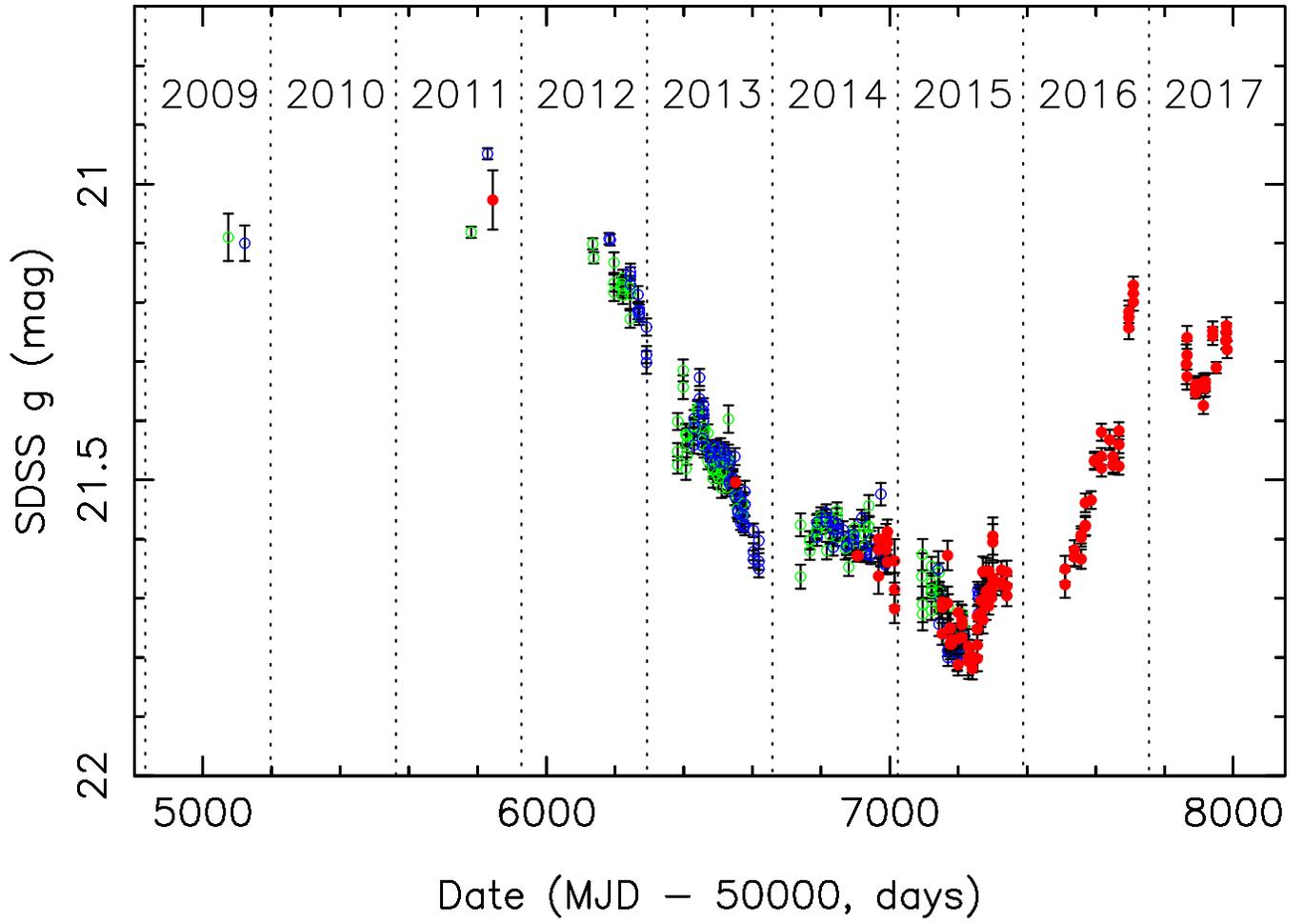}
\caption[.]  {The combined A (open blue circles) $+$ B (open green circles) light curve, together with the C (filled red circles) light curve, the latter
shifted by $\tau_{\rm CA} = \ACshort$\,days and $\Delta m_{\rm AC} = \ACshortoffset$.  The dates and magnitude values correspond to the light curve of the A image. }
\label{fig:addC}
\end{figure*}

As shown in Figure~\ref{fig:addC}, the best-fit value derived for $\tau_{\rm CA}$ makes very distinctive predictions for the photometric behaviour of images A and B during the 2015 and 2016 monitoring seasons: The A and B light curves are both predicted to reach a sharp minimum, spaced by their \shortABdisp - day time delay, in mid-2015.  
This minimum corresponds to the lowest luminosity state of the quasar observed during the $\sim 8$\, years of intrinsic quasar brightness variations spanned by all available photometric data. Our most recent data points, from 2015 Aug. 22, 24, and 26 UT (see Table~\ref{tab:photometrytable}), 
show an increase in the flux of image A by $\sim 0.1$ magnitudes after passing the global minimum in the intrinsic quasar light curve, exactly as predicted by the best-fit $\tau_{\rm CA}$ value. Continued monitoring of images A and B, tracking their predicted further brightening during 
the fall of 2015, should substantially reduce the size of the uncertainty intervals of both $\tau_{\rm AB}$ and $\tau_{\rm CA}$.   

The fluxes of A and B are predicted to further increase by $> 0.5$\, magnitude during 2016. Observing these predicted variations of image A and B through the end of 2016 will strictly exclude any alternative time delay values in the broad interval $830 \lesssim \tau_{\rm CA} \lesssim 2100$\, days. 
We note that the upper limit of this interval extends well beyond the range of time delays allowed by the lens model of Dahle et al.\ (2013).    



\section{Summary and discussion} 
\label{sec:discuss}

Predictions from the lens model presented by Dahle et al.\  (2013) indicate that three independent 
time delays $\tau_{\rm AB}$, $\tau_{\rm ED}$ and $\tau_{\rm FE}$ in the system are all of order 100 days. 
The predicted time delays between image C and the pair (A-B) and between (A-B) and the central images (D-E-F) 
are of order 1300 days and 700-1000 days, respectively (with $\sim 40$\% uncertainty). 

We have measured $\tau_{\rm AB}$ using two different methods and find results which are consistent at the $2\sigma$ level. 
However, unlike the $\chi^2$ minimisation method, the dispersion method does not make any assumptions about the photometric 
behaviour of the quasar images during the $\sim 14$\,day sampling intervals, and we therefore adopt the dispersion method value of 
$\tau_{\rm AB} = \ABdisp$\,days (95\% confidence interval) as the most robust measurement. This is consistent 
with the model predicted value of $\tau_{\rm AB} = 112^{+225}_{-158}$\, days. 
The measured magnitude difference $\Delta m_{\rm AB} = \ABdispoffset$ is also consistent 
with the ratio of the model-predicted magnifications of $\mu_A = 5.4^{+4.6}_{-1.6}$ and $\mu_B = 4.7^{+3.4}_{-1.6}$ (Dahle et al.\ 2013). 

The measured $\tau_{\rm AB}$ time delay is very close to 1.5 synodic months, implying that it is possible to construct a continuous (apart from seasonal gaps) quasar light curve from the combination of data points from image A and image B which will be minimally affected by poor-$S/N$ measurements caused by bright moonlight. 

From our systematic monitoring campaign, there is so far only a year of overlap between the intrinsic quasar 
light curve derived from image C and those from A and B. However, the strong amplitude of the intrinsic brightness fluctuations of the quasar, 
the presence of earlier data points from 2009 and 2011, and the excellent match of the sharp flux minima observed in image C in 2013 
and in images A and B in 2015 enable us to measure the  $\tau_{\rm CA}$ time delay from the data set currently in hand. 

The best-fit value of  $\tau_{\rm CA} = \ACdisp$\,days is in slight tension with the model-predicted value of $\tau_{\rm CA} = 1256^{+485}_{-411}$\, days of Dahle et al.\ (2013). We note that ongoing work, refining the lens model by incorporating additional constraints from newly identified 
multiply imaged source galaxies in this lens system, produce systematically lower model-predicted values for $\tau_{\rm CA}$.  
The measured magnitude difference $\Delta m_{\rm AC} = \ACdispoffset$ is  consistent within the errors with the ratio of the model-predicted magnifications of $\mu_A = 5.4^{+4.6}_{-1.6}$ and $\mu_C = 2.4^{+1.1}_{-0.3}$ (Dahle et al.\ 2013). 

Our prediction of the timing of the expected $0.5-0.6$\,magnitude rise in images A and B 
can be used to plan a more intensive campaign to monitor photometric and spectroscopic changes in the quasar 
during this phase in 2016. This would e.g., allow a reverberation mapping study of the $z=2.82$ quasar. 

A subsequent corresponding brightness increase should occur for the three fainter images around 2018, given the Dahle et al.\  (2013) 
prediction for the time delays between the (A-B) pair and the (D-F) trio.   
During 2015, these fainter images should be entering the decline phase seen in images A and B in 2012-2013, before reaching the plateau 
seen in 2014. Provided that even rough estimates of the time delays can be made from observations in the decline and subsequent plateau phase, 
a targeted campaign at the predicted time of the subsequent rise could provide accurate time delay estimates for the 
three demagnified images in the cluster center. 

We have recently obtained 5-band HST optical/NIR imaging (GO-13337; PI: Sharon) of \QSO \, which confirms the existence of the sixth quasar image and reveals additional multiply lensed 
background galaxies. Additional constraints from the positions and redshifts of these, together with the measurements reported 
here for  $\tau_{\rm AB} $ and $\tau_{\rm CA} $, will allow us to refine the lens model of this system and provide updated predictions for the unmeasured 
time delays.  The HST data will also be crucial for modelling the brightness distribution of the brightest galaxies in the cluster core and 
removing their contribution to the measured flux of images D-F in ground-based photometric monitoring data.  

To summarise: The steep quasar brightness fluctuations evident from the light curve of the leading C image provide strong constraining power   
to measurements of the time delays between all six known images in this lens system. 
The range of clustercentric radii over which such measurements are possible is truly unique for this system. 
The next 2-3 observing seasons will be instrumental in pinning down these time delays, enabling a wide 
range of physical studies of both the cluster lens and the quasar source.

\acknowledgments 

HD acknowledges support from the Research Council of Norway. MBB acknowledges support from the NSF through grant AST-1009012.
We acknowledge use of the new Gemini Fast Turnaround Program and the Fast-Track Service mode at NOT, which both provided 
important data for this paper. We thank the staffs of NOT and Gemini for their flexibility in supporting the execution of our monitoring campaign. 
The authors wish to recognize and acknowledge the very significant cultural role and reverence that the summit of Mauna Kea has always had within the indigenous Hawaiian community.  We are most fortunate to have the opportunity to conduct observations from this mountain.

\clearpage
\LongTables
\begin{deluxetable}{cccccc}
\tablewidth{0pt}
\tablecaption{$g$-band photometry of the three brightest quasar images in SDSS J2222+2745 }
\tablehead{
\colhead{MJD}           &
\colhead{Image A}       & \colhead{Image B}  &
\colhead{Image C}       &
\colhead{Observatory} & \colhead{Detector}}
\startdata
 5122.200 & 21.100$\pm$ 0.030 & 21.430$\pm$ 0.040 & 21.510$\pm$ 0.050 &     SDSS &     SDSS  \cr 
 5828.935 & 20.949$\pm$ 0.009 & 21.422$\pm$ 0.009 & 21.987$\pm$ 0.009 &      NOT &    MOSCA  \cr 
 6182.963 & 21.092$\pm$ 0.009 & 21.441$\pm$ 0.009 & 22.111$\pm$ 0.009 &      NOT &    MOSCA  \cr 
 6185.985 & 21.094$\pm$ 0.009 & 21.465$\pm$ 0.009 & 22.111$\pm$ 0.009 &      NOT &    MOSCA  \cr 
 6244.907 & 21.170$\pm$ 0.014 & 21.473$\pm$ 0.017 & 22.145$\pm$ 0.030 &      NOT &   ALFOSC  \cr 
 6244.911 & 21.148$\pm$ 0.014 & 21.524$\pm$ 0.014 & 22.083$\pm$ 0.019 &      NOT &   ALFOSC  \cr 
 6244.915 & 21.155$\pm$ 0.014 & 21.507$\pm$ 0.014 & 22.100$\pm$ 0.021 &      NOT &   ALFOSC  \cr 
 6266.907 & 21.187$\pm$ 0.014 & 21.509$\pm$ 0.015 & 22.101$\pm$ 0.023 &      NOT &   ALFOSC  \cr 
 6266.911 & 21.214$\pm$ 0.014 & 21.513$\pm$ 0.015 & 22.080$\pm$ 0.022 &      NOT &   ALFOSC  \cr 
 6270.830 & 21.218$\pm$ 0.015 & 21.529$\pm$ 0.014 & 22.122$\pm$ 0.021 &      NOT &   ALFOSC  \cr 
 6270.834 & 21.213$\pm$ 0.015 & 21.518$\pm$ 0.014 & 22.089$\pm$ 0.020 &      NOT &   ALFOSC  \cr 
 6270.838 & 21.224$\pm$ 0.015 & 21.499$\pm$ 0.014 & 22.070$\pm$ 0.020 &      NOT &   ALFOSC  \cr 
 6291.811 & 21.301$\pm$ 0.019 & 21.527$\pm$ 0.024 & 22.120$\pm$ 0.037 &      NOT &   ALFOSC  \cr 
 6291.815 & 21.242$\pm$ 0.015 & 21.516$\pm$ 0.017 & 22.168$\pm$ 0.028 &      NOT &   ALFOSC  \cr 
 6291.819 & 21.288$\pm$ 0.014 & 21.568$\pm$ 0.015 & 22.201$\pm$ 0.024 &      NOT &   ALFOSC  \cr 
 6430.182 & 21.396$\pm$ 0.015 & 21.792$\pm$ 0.015 & 22.189$\pm$ 0.019 &      NOT &   ALFOSC  \cr 
 6430.189 & 21.441$\pm$ 0.015 & 21.815$\pm$ 0.015 & 22.243$\pm$ 0.019 &      NOT &   ALFOSC  \cr 
 6430.197 & 21.413$\pm$ 0.015 & 21.741$\pm$ 0.014 & 22.198$\pm$ 0.019 &      NOT &   ALFOSC  \cr 
 6446.168 & 21.363$\pm$ 0.015 & 21.683$\pm$ 0.020 & 22.191$\pm$ 0.028 &      NOT &   ALFOSC  \cr 
 6446.182 & 21.327$\pm$ 0.014 & 21.655$\pm$ 0.019 & 22.111$\pm$ 0.025 &      NOT &   ALFOSC  \cr 
 6454.178 & 21.430$\pm$ 0.014 & 21.821$\pm$ 0.019 & 22.234$\pm$ 0.026 &      NOT &   ALFOSC  \cr 
 6454.182 & 21.443$\pm$ 0.015 & 21.763$\pm$ 0.018 & 22.261$\pm$ 0.026 &      NOT &   ALFOSC  \cr 
 6454.186 & 21.382$\pm$ 0.014 & 21.764$\pm$ 0.018 & 22.233$\pm$ 0.025 &      NOT &   ALFOSC  \cr 
 6457.183 & 21.388$\pm$ 0.007 & 21.767$\pm$ 0.014 & 22.254$\pm$ 0.015 &      NOT &   ALFOSC  \cr 
 6457.190 & 21.401$\pm$ 0.011 & 21.800$\pm$ 0.014 & 22.260$\pm$ 0.016 &      NOT &   ALFOSC  \cr 
 6457.198 & 21.373$\pm$ 0.011 & 21.778$\pm$ 0.014 & 22.260$\pm$ 0.016 &      NOT &   ALFOSC  \cr 
 6477.090 & 21.452$\pm$ 0.015 & 21.765$\pm$ 0.014 & 22.207$\pm$ 0.019 &      NOT &   ALFOSC  \cr 
 6477.097 & 21.455$\pm$ 0.014 & 21.749$\pm$ 0.014 & 22.251$\pm$ 0.018 &      NOT &   ALFOSC  \cr 
 6477.105 & 21.463$\pm$ 0.014 & 21.767$\pm$ 0.014 & 22.295$\pm$ 0.018 &      NOT &   ALFOSC  \cr 
 6488.130 & 21.444$\pm$ 0.010 & 21.721$\pm$ 0.015 & 22.220$\pm$ 0.016 &      NOT &   ALFOSC  \cr 
 6488.137 & 21.472$\pm$ 0.010 & 21.728$\pm$ 0.015 & 22.249$\pm$ 0.016 &      NOT &   ALFOSC  \cr 
 6488.145 & 21.443$\pm$ 0.010 & 21.719$\pm$ 0.015 & 22.227$\pm$ 0.016 &      NOT &   ALFOSC  \cr 
 6507.150 & 21.470$\pm$ 0.016 & 21.781$\pm$ 0.021 & 22.289$\pm$ 0.030 &      NOT &   ALFOSC  \cr 
 6507.158 & 21.455$\pm$ 0.016 & 21.757$\pm$ 0.021 & 22.265$\pm$ 0.030 &      NOT &   ALFOSC  \cr 
 6507.165 & 21.445$\pm$ 0.015 & 21.751$\pm$ 0.020 & 22.283$\pm$ 0.030 &      NOT &   ALFOSC  \cr 
 6518.141 & 21.451$\pm$ 0.015 & 21.761$\pm$ 0.015 & 22.288$\pm$ 0.020 &      NOT &   ALFOSC  \cr 
 6518.148 & 21.471$\pm$ 0.014 & 21.808$\pm$ 0.014 & 22.302$\pm$ 0.018 &      NOT &   ALFOSC  \cr 
 6518.156 & 21.459$\pm$ 0.013 & 21.789$\pm$ 0.014 & 22.281$\pm$ 0.017 &      NOT &   ALFOSC  \cr 
 6533.039 & 21.475$\pm$ 0.015 & 21.812$\pm$ 0.014 & 22.213$\pm$ 0.017 &      NOT &   ALFOSC  \cr 
 6533.050 & 21.502$\pm$ 0.015 & 21.824$\pm$ 0.015 & 22.262$\pm$ 0.019 &      NOT &   ALFOSC  \cr 
 6533.058 & 21.507$\pm$ 0.015 & 21.820$\pm$ 0.015 & 22.235$\pm$ 0.019 &      NOT &   ALFOSC  \cr 
 6533.065 & 21.466$\pm$ 0.014 & 21.837$\pm$ 0.015 & 22.285$\pm$ 0.022 &      NOT &   ALFOSC  \cr 
 6543.103 & 21.497$\pm$ 0.014 & 21.799$\pm$ 0.015 & 22.212$\pm$ 0.020 &      NOT &   ALFOSC  \cr 
 6543.110 & 21.496$\pm$ 0.014 & 21.830$\pm$ 0.015 & 22.208$\pm$ 0.019 &      NOT &   ALFOSC  \cr 
 6543.118 & 21.501$\pm$ 0.014 & 21.788$\pm$ 0.015 & 22.187$\pm$ 0.019 &      NOT &   ALFOSC  \cr 
 6549.047 & 21.528$\pm$ 0.015 & 21.836$\pm$ 0.020 & 22.138$\pm$ 0.026 &      NOT &   ALFOSC  \cr 
 6549.054 & 21.461$\pm$ 0.014 & 21.815$\pm$ 0.017 & 22.220$\pm$ 0.023 &      NOT &   ALFOSC  \cr 
 6557.850 & 21.555$\pm$ 0.014 & 21.811$\pm$ 0.017 & 22.197$\pm$ 0.023 &      NOT &   ALFOSC  \cr 
 6557.854 & 21.552$\pm$ 0.014 & 21.853$\pm$ 0.017 & 22.137$\pm$ 0.022 &      NOT &   ALFOSC  \cr 
 6557.858 & 21.523$\pm$ 0.015 & 21.826$\pm$ 0.018 & 22.172$\pm$ 0.024 &      NOT &   ALFOSC  \cr 
 6565.970 & 21.573$\pm$ 0.012 & 21.800$\pm$ 0.015 & 22.169$\pm$ 0.014 &      NOT &   ALFOSC  \cr 
 6565.977 & 21.559$\pm$ 0.013 & 21.829$\pm$ 0.014 & 22.196$\pm$ 0.015 &      NOT &   ALFOSC  \cr 
 6565.985 & 21.531$\pm$ 0.011 & 21.811$\pm$ 0.015 & 22.137$\pm$ 0.014 &      NOT &   ALFOSC  \cr 
 6573.944 & 21.579$\pm$ 0.010 & 21.810$\pm$ 0.015 & 22.162$\pm$ 0.014 &      NOT &   ALFOSC  \cr 
 6573.951 & 21.560$\pm$ 0.010 & 21.806$\pm$ 0.015 & 22.182$\pm$ 0.014 &      NOT &   ALFOSC  \cr 
 6573.958 & 21.579$\pm$ 0.009 & 21.853$\pm$ 0.015 & 22.170$\pm$ 0.014 &      NOT &   ALFOSC  \cr 
 6578.051 & 21.520$\pm$ 0.018 & 21.737$\pm$ 0.023 & 22.088$\pm$ 0.031 &      NOT &   ALFOSC  \cr 
 6578.055 & 21.542$\pm$ 0.020 & 21.806$\pm$ 0.025 & 22.078$\pm$ 0.031 &      NOT &   ALFOSC  \cr 
 6578.059 & 21.577$\pm$ 0.017 & 21.841$\pm$ 0.022 & 22.149$\pm$ 0.027 &      NOT &   ALFOSC  \cr 
 6602.891 & 21.634$\pm$ 0.015 & 21.868$\pm$ 0.014 & 22.135$\pm$ 0.015 &      NOT &   ALFOSC  \cr 
 6602.899 & 21.621$\pm$ 0.014 & 21.889$\pm$ 0.015 & 22.160$\pm$ 0.018 &      NOT &   ALFOSC  \cr 
 6602.906 & 21.585$\pm$ 0.011 & 21.862$\pm$ 0.015 & 22.156$\pm$ 0.014 &      NOT &   ALFOSC  \cr 
 6618.884 & 21.650$\pm$ 0.014 & 21.899$\pm$ 0.015 & 22.179$\pm$ 0.018 &      NOT &   ALFOSC  \cr 
 6618.888 & 21.638$\pm$ 0.014 & 21.880$\pm$ 0.015 & 22.163$\pm$ 0.017 &      NOT &   ALFOSC  \cr 
 6618.891 & 21.602$\pm$ 0.014 & 21.903$\pm$ 0.016 & 22.139$\pm$ 0.020 &      NOT &   ALFOSC  \cr 
 6788.186 & 21.572$\pm$ 0.015 & 21.916$\pm$ 0.019 & 22.160$\pm$ 0.022 &      NOT &   ALFOSC  \cr 
 6788.193 & 21.597$\pm$ 0.015 & 22.003$\pm$ 0.020 & 22.133$\pm$ 0.022 &      NOT &   ALFOSC  \cr 
 6815.173 & 21.556$\pm$ 0.015 & 21.942$\pm$ 0.015 & 22.113$\pm$ 0.016 &      NOT &   ALFOSC  \cr 
 6815.181 & 21.570$\pm$ 0.015 & 21.960$\pm$ 0.015 & 22.112$\pm$ 0.016 &      NOT &   ALFOSC  \cr 
 6815.188 & 21.565$\pm$ 0.015 & 21.942$\pm$ 0.015 & 22.101$\pm$ 0.016 &      NOT &   ALFOSC  \cr 
 6835.176 & 21.614$\pm$ 0.014 & 21.927$\pm$ 0.014 & 22.076$\pm$ 0.015 &      NOT &   ALFOSC  \cr 
 6835.184 & 21.584$\pm$ 0.015 & 21.914$\pm$ 0.014 & 22.080$\pm$ 0.015 &      NOT &   ALFOSC  \cr 
 6835.191 & 21.571$\pm$ 0.015 & 21.924$\pm$ 0.015 & 22.117$\pm$ 0.016 &      NOT &   ALFOSC  \cr 
 6847.162 & 21.584$\pm$ 0.014 & 21.901$\pm$ 0.015 & 22.022$\pm$ 0.016 &      NOT &   ALFOSC  \cr 
 6847.169 & 21.574$\pm$ 0.014 & 21.916$\pm$ 0.015 & 22.062$\pm$ 0.016 &      NOT &   ALFOSC  \cr 
 6847.179 & 21.589$\pm$ 0.014 & 21.908$\pm$ 0.016 & 22.060$\pm$ 0.016 &      NOT &   ALFOSC  \cr 
 6864.105 & 21.584$\pm$ 0.015 & 21.959$\pm$ 0.015 & 22.017$\pm$ 0.015 &      NOT &   ALFOSC  \cr 
 6874.056 & 21.610$\pm$ 0.014 & 21.914$\pm$ 0.015 & 21.950$\pm$ 0.015 &      NOT &   ALFOSC  \cr 
 6874.063 & 21.612$\pm$ 0.015 & 21.934$\pm$ 0.014 & 21.952$\pm$ 0.014 &      NOT &   ALFOSC  \cr 
 6894.020 & 21.610$\pm$ 0.012 & 21.937$\pm$ 0.014 & 21.963$\pm$ 0.014 &      NOT &   ALFOSC  \cr 
 6894.028 & 21.592$\pm$ 0.014 & 21.902$\pm$ 0.014 & 21.903$\pm$ 0.014 &      NOT &   ALFOSC  \cr 
 6894.035 & 21.599$\pm$ 0.015 & 21.892$\pm$ 0.014 & 21.943$\pm$ 0.014 &      NOT &   ALFOSC  \cr 
 6918.096 & 21.565$\pm$ 0.014 & 21.956$\pm$ 0.018 & 21.915$\pm$ 0.017 &      NOT &   ALFOSC  \cr 
 6928.011 & 21.619$\pm$ 0.015 & 21.952$\pm$ 0.014 & 21.956$\pm$ 0.014 &      NOT &   ALFOSC  \cr 
 6928.019 & 21.620$\pm$ 0.014 & 21.987$\pm$ 0.015 & 21.944$\pm$ 0.015 &      NOT &   ALFOSC  \cr 
 6928.026 & 21.625$\pm$ 0.013 & 21.942$\pm$ 0.014 & 21.959$\pm$ 0.014 &      NOT &   ALFOSC  \cr 
 6944.979 & 21.600$\pm$ 0.014 & 21.933$\pm$ 0.015 & 21.923$\pm$ 0.014 &      NOT &   ALFOSC  \cr 
 6944.986 & 21.628$\pm$ 0.015 & 21.921$\pm$ 0.014 & 21.960$\pm$ 0.014 &      NOT &   ALFOSC  \cr 
 6944.993 & 21.625$\pm$ 0.014 & 21.933$\pm$ 0.015 & 21.900$\pm$ 0.015 &      NOT &   ALFOSC  \cr 
 6973.943 & 21.629$\pm$ 0.018 & 21.931$\pm$ 0.024 & 21.699$\pm$ 0.019 &      NOT &   ALFOSC  \cr 
 6973.950 & 21.627$\pm$ 0.018 & 21.920$\pm$ 0.024 & 21.726$\pm$ 0.019 &      NOT &   ALFOSC  \cr 
 6973.958 & 21.524$\pm$ 0.019 & 21.952$\pm$ 0.027 & 21.709$\pm$ 0.021 &      NOT &   ALFOSC  \cr 
 6986.896 & 21.643$\pm$ 0.015 & 21.883$\pm$ 0.018 & 21.654$\pm$ 0.015 &      NOT &   ALFOSC  \cr 
 6986.903 & 21.642$\pm$ 0.014 & 21.920$\pm$ 0.017 & 21.683$\pm$ 0.014 &      NOT &   ALFOSC  \cr 
 6986.911 & 21.623$\pm$ 0.014 & 21.918$\pm$ 0.017 & 21.668$\pm$ 0.014 &      NOT &   ALFOSC  \cr 
 7141.226 & 21.651$\pm$ 0.031 & 22.003$\pm$ 0.039 & 21.788$\pm$ 0.034 &      NOT &   ALFOSC  \cr 
 7143.207 & 21.696$\pm$ 0.024 & 22.050$\pm$ 0.030 & 21.772$\pm$ 0.024 &      NOT &   ALFOSC  \cr 
 7143.211 & 21.744$\pm$ 0.020 & 22.067$\pm$ 0.027 & 21.809$\pm$ 0.022 &      NOT &   ALFOSC  \cr 
 7143.214 & 21.696$\pm$ 0.018 & 21.966$\pm$ 0.026 & 21.742$\pm$ 0.019 &      NOT &   ALFOSC  \cr 
 7168.539 & 21.789$\pm$ 0.009 & 22.031$\pm$ 0.009 & 21.836$\pm$ 0.009 &   Gemini &     GMOS  \cr 
 7169.188 & 21.750$\pm$ 0.014 & 21.986$\pm$ 0.016 & 21.822$\pm$ 0.014 &      NOT &   ALFOSC  \cr 
 7169.195 & 21.775$\pm$ 0.014 & 22.061$\pm$ 0.016 & 21.824$\pm$ 0.014 &      NOT &   ALFOSC  \cr 
 7169.203 & 21.800$\pm$ 0.014 & 22.014$\pm$ 0.017 & 21.826$\pm$ 0.014 &      NOT &   ALFOSC  \cr 
 7169.518 & 21.787$\pm$ 0.009 & 22.025$\pm$ 0.009 & 21.830$\pm$ 0.009 &   Gemini &     GMOS  \cr 
 7191.154 & 21.771$\pm$ 0.014 & 22.044$\pm$ 0.015 & 21.818$\pm$ 0.014 &      NOT &   ALFOSC  \cr 
 7191.161 & 21.777$\pm$ 0.014 & 22.029$\pm$ 0.016 & 21.822$\pm$ 0.014 &      NOT &   ALFOSC  \cr 
 7191.168 & 21.783$\pm$ 0.014 & 21.996$\pm$ 0.016 & 21.857$\pm$ 0.014 &      NOT &   ALFOSC  \cr 
 7196.183 & 21.784$\pm$ 0.015 & 22.052$\pm$ 0.014 & 21.828$\pm$ 0.014 &      NOT &   ALFOSC  \cr 
 7196.190 & 21.809$\pm$ 0.014 & 22.039$\pm$ 0.015 & 21.817$\pm$ 0.014 &      NOT &   ALFOSC  \cr 
 7196.197 & 21.807$\pm$ 0.014 & 22.059$\pm$ 0.015 & 21.819$\pm$ 0.014 &      NOT &   ALFOSC  \cr 
 7218.080 & 21.794$\pm$ 0.016 & 22.110$\pm$ 0.022 & 21.740$\pm$ 0.015 &      NOT &   ALFOSC  \cr 
 7218.087 & 21.763$\pm$ 0.016 & 22.054$\pm$ 0.021 & 21.731$\pm$ 0.016 &      NOT &   ALFOSC  \cr 
 7228.476 & 21.791$\pm$ 0.009 & 22.096$\pm$ 0.009 & 21.793$\pm$ 0.009 &   Gemini &     GMOS  \cr 
 7256.088 & 21.687$\pm$ 0.014 & 22.066$\pm$ 0.016 & 21.748$\pm$ 0.014 &      NOT &   ALFOSC  \cr 
 7258.077 & 21.725$\pm$ 0.014 & 22.079$\pm$ 0.014 & 21.746$\pm$ 0.014 &      NOT &   ALFOSC  \cr 
 7258.084 & 21.697$\pm$ 0.015 & 22.121$\pm$ 0.015 & 21.723$\pm$ 0.015 &      NOT &   ALFOSC  \cr 
 7258.091 & 21.696$\pm$ 0.015 & 22.066$\pm$ 0.015 & 21.733$\pm$ 0.015 &      NOT &   ALFOSC  \cr 
 7260.148 & 21.691$\pm$ 0.014 & 22.087$\pm$ 0.017 & 21.763$\pm$ 0.014 &      NOT &   ALFOSC  \cr 
\tableline
\enddata
\tablecomments{The Modified Julian Days (MJD) column gives the date of the observation
  relative to MJD$=50000$.     
  } 
\label{tab:photometrytable}  
\end{deluxetable}

\end{document}